\documentclass[12pt]{article}
\usepackage{latexsym}
\usepackage{epsf}

\def\beq{\begin{equation}}
\def\eeq{\end{equation}}
\def\bqry{\begin{eqnarray}}
\def\eqry{\end{eqnarray}}
\def\NON{\nonumber\\}
\def\NXT{\\}
\def\lbl{\label}
\def\bibi{\bibitem}

\def\a{\alpha}
\def\b{\beta}
\def\c{\chi}
\def\d{\delta}
\def\e{\epsilon}                
\def\g{\gamma}
\def\h{\eta}

\def\j{\psi}

\def\l{\lambda}
\def\m{\mu}
\def\n{\nu}

\def\p{\pi}                     
\def\th{\theta}                  
\def\r{\rho}                    
\def\s{\sigma}                  
\def\t{\tau}

\def\x{\xi}

\def\F{\Phi}
\def\G{\Gamma}
\def\J{\Psi}

\def\S{\Sigma}

\def\cc{{\cal C}}
\def\cd{{\cal D}}

\def\cg{{\cal G}}

\def\cl{{\cal L}}
\def\cm{{\cal M}}
\def\cn{{\cal N}}
\def\co{{\cal O}}
\def\cp{{\cal P}}

\def\cs{{\cal S}}

\def\cbo{{\,\raise-.15ex\Sc [\,}}                       

\def\Sl#1{\rlap{\hbox{$\mskip 3 mu /$}}#1}      
\def\vev#1{\Big\langle #1 \Big\rangle}           

\def\sbra#1{\left\langle #1\right|}             
\def\sket#1{\left| #1\right\rangle}             
\def\svev#1{\left\langle #1\right\rangle}       

\def\lvec#1{\raisebox{0.0ex}{$\stackrel{\leftarrow}{#1}$} }  

\def\ddt#1{{\buildrel {\hbox{\LARGE .\kern-2pt.}} \over {#1}}}

\def\secteq#1{ \setcounter{equation}{0}
               \renewcommand{\theequation}{#1.\arabic{equation}} }

\def\beqn#1{ \renewcommand{\theequation}{#1}
             \begin{eqnarray} }
\def\eeqn{ \renewcommand{\theequation}{\arabic{equation}}
           \end{eqnarray} }

\def\beqr#1{ \setcounter{equation}{#1}
             \begin{eqnarray} }

\def\eeqr{\end{eqnarray}}

\def\beqrabc#1{ \setcounter{equation}{0}
                \renewcommand{\theequation}{#1\alph{equation}}
                \begin{eqnarray} }
\def\beqrn#1#2{ \setcounter{equation}{#2}
                \renewcommand{\theequation}{#1.\arabic{equation}}
                \begin{eqnarray} }

\def\seeq#1{eq.~(\ref{#1})}
\def\seEq#1{Eq.~(\ref{#1})}
\def\seeqs#1{eqs.~(\ref{#1})}

\def\seneq#1{~(\ref{#1})}

\def\rcite#1{ref.~\cite{#1}}
\def\rcites#1{refs.~\cite{#1}}

\def\NPB#1{Nucl. Phys. {\bf B#1}}
\def\NPBP#1{Nucl. Phys. (Proc. Suppl.) {\bf B#1}}
\def\PLB#1{Phys. Lett. {\bf B#1}}
\def\PRD#1{Phys. Rev. {\bf D#1}}
\def\PR#1{Phys. Rev. {\bf #1}}
\def\PRL#1{Phys. Rev. Lett. {\bf #1}}
\def\PRP#1{Phys. Rep. {\bf #1}}

\def\ssstyle{\scriptscriptstyle}
\def\sstyle{\scriptstyle}

\def\ie{\mbox{\it i.e.} }
\def\eg{\mbox{\it e.g.} }

\def\leqx{\,\raisebox{-1.0ex}{$\stackrel{\textstyle <}{\sim}$}\,}

\def\frac#1#2{ {\sstyle {#1\over #2} } }
\def\det#1{{\rm det}\left(#1\right)}

\def\sbraket#1#2{\left\langle#1|#2\right\rangle} 

\def\tr{{\rm tr}\,}

\def\half{{1\over 2}}

\def\det{{\rm det\,}}
\def\Det{{\rm Det}}

\thicklines                         
\def\lvec#1{{\buildrel \leftarrow\over {#1}}}

\def\pd{\partial}
\def\hp{\hat{\partial}}
\def\hd{\hat{D}}
\def\bJ{\overline\Psi}
\def\bj{\overline\psi}
\def\bc{\overline\chi}
\def\bji{\overline\psi{}^i}
\def\bci{\overline\chi{}^i}
\def\jl{J^L}
\def\hjl{\hat{J}^L}
\def\jlt{J^{\rm latt}}
\def\tjlt{\tilde{J}^{\rm latt}}
\def\jllt{J^{\rm L,latt}}
\def\tjllt{\tilde{J}^{\rm L,latt}}
\def\eg{e.g.\ }

\def\sixteen{\hspace{-.5ex} \mbox{\boldmath $16$} \hspace{-.5ex}}
\def\tenrep{\hspace{-1ex} \mbox{\boldmath $10$} \hspace{-.0ex}}
\def\sixbar{\hspace{-.5ex} \mbox{\boldmath $\overline{16}$} \hspace{-.5ex}}
\def\idn{\hspace{.5ex}I\hspace{.5ex}}

\def\hb{\raisebox{.2ex}{$\hat{\raisebox{-.6ex}{$\Box$}}$}}

\begin{document}
\hyphenation{fer-mio-nic per-tur-ba-tive pa-ra-me-tri-za-tion
pa-ra-me-tri-zed a-nom-al-ous}

\hfill TAUP--2697--02 \\

\begin{center}
\vspace{15mm}
{\large\bf Fermion-number violation in regularizations\\
\vspace{1ex} that preserve fermion-number symmetry}
\\[15mm]
Maarten Golterman$^a$\ \ and \ \ Yigal Shamir$^b$
\\[10mm]
{\small\it
$^a$Department of Physics and Astronomy,
San Francisco State University\\
San Francisco, CA 94132, USA}\\
{\tt maarten@quark.sfsu.edu}
\\[5mm]
{\small\it $^b$School of Physics and Astronomy\\
Tel-Aviv University, Ramat~Aviv,~69978~ISRAEL}\\
{\tt shamir@post.tau.ac.il}
\\[10mm]
{ABSTRACT}
\\[2mm]
\end{center}

\begin{quotation}
There exist both continuum and lattice regularizations of gauge theories
with fermions
which preserve chiral U(1) invariance (``fermion number").
Such regularizations necessarily break
gauge invariance but, in a covariant
gauge,  one recovers gauge invariance
to all orders in perturbation theory by including suitable counter terms.

At the non-perturbative level, an apparent conflict then arises
between the chiral U(1) symmetry of the regularized theory and the existence
of 't~Hooft vertices in the renormalized theory.
The only possible resolution of the paradox is that the chiral U(1) symmetry
is broken spontaneously in the enlarged Hilbert space of
the covariantly gauge-fixed theory. The corresponding Goldstone pole
is unphysical. The theory must therefore be defined
by introducing a small fermion-mass term that breaks explicitly
the chiral U(1) invariance, and is sent to zero
after the infinite-volume limit has been taken.
Using this careful definition (and a lattice regularization)
for the calculation of correlation functions in the one-instanton sector,
we show that the 't~Hooft vertices are recovered as expected.
\end{quotation}

\newpage
\vspace{5ex}
\noindent {\large\bf 1.~Introduction and Conclusion}
\secteq{1}
\vspace{3ex}

Every gauge theory coupled to massless fermions
has an anomalous chiral current. Representing all fermions by
left-handed Weyl fields, the Noether current associated
with a common global U(1) rotation is classically conserved.
At the one-loop level, a gauge-invariant definition of the current yields
the Adler-Bell-Jackiw anomaly~\cite{abj}
\beq
  \partial_\m \jl_\m = {c g^2\over 8 \p^2}\; \tr F\tilde{F} \,.
\lbl{dj}
\eeq
For our notation see Appendix~A.
The group-theoretical constant $c$ is additive.
(Each Weyl fermion in the fundamental representation contributes $c=\half$.
We will assume the gauge symmetry to be non-anomalous throughout this paper.)
One can also define a conserved but gauge non-invariant current
\beq
  \hjl_\m = \jl_\m - g^2\,K_\m \,,
\lbl{hatj}
\eeq
where
\beq
  K_\m = {c\over 8 \p^2}\; \e_{\m\n\r\s}\,
         \tr \!\left(A_\n F_{\r\s} - {1\over 3}A_\n A_\r A_\s \right) \,.
\lbl{K}
\eeq
If a gauge-invariant regularization is used,
the gauge-invariant, non-conserved current $\jl_\m$ is defined
(up to a $Z$ factor) by the fermion bilinear
\beq
  \sum_i \bji_L \,\s_\m\, \j_L^i\,.
\lbl{jl}
\eeq
Here $i$ runs over all the left-handed fields.
In QCD-like theories, this applies in particular
to dimensional regularization as well as to
the standard lattice regularization~\cite{W,KS}.

What happens if the regulator is chiral-U(1) invariant?
The U(1) current will now be conserved at the one-loop level.
Therefore it must, when the cutoff is removed, coincide
with the gauge non-invariant current
$\hjl_\m$ defined in \seeq{hatj}. (This is true up to a term
$\partial_\nu H_{\mu\nu}$ with $H_{\mu\nu}$ an anti-symmetric tensor.)
Since, classically, the U(1) Noether current is gauge invariant,
this can only happen because the regularization itself is not gauge invariant:
a chiral-U(1)-invariant regularization is, necessarily, not gauge invariant.

Does this observation imply that all chiral-U(1)-invariant
regularizations must be dismissed? To begin with, in perturbation theory
the answer is no, provided the action
contains covariant gauge-fixing (and ghost) terms.
A covariant gauge will be assumed throughout this paper.
In the presence of a longitudinal kinetic term, $(\partial_\m A_\m)^2$,
the theory is renormalizable by power-counting {\it without}
relying on gauge invariance. The renormalization program
reduces to an algebraic problem
and (provided the gauge symmetry is non-anomalous)
one can restore gauge invariance to all orders in perturbation
theory by suitable counter terms (see \eg \rcite{algrnr}).

Beyond perturbation theory, there is an apparent conflict
between chiral U(1) invariance of the regularized theory,
and the fact that instanton-mediated amplitudes violate
the conservation of the chiral U(1) charge~\cite{thft}.
It has been pointed out long ago~\cite{KgSs,SC,thftrev} that,
in a covariant gauge, the breaking of chiral U(1) invariance can be
{\it spontaneous} in a technical sense. The reason is that the
enlarged Hilbert space of the gauge-fixed theory can accommodate
a new Goldstone pole. The latter is unphysical since it originates from
the $K_\m$ part of the current
$\hjl_\m$. If the regulator is chiral-U(1) invariant
there is, in fact, no other possibility. In this paper,
we re-examine this question in the context of a specific
lattice-regularization method.
Our analysis reveals that a careful
definition of the thermodynamical limit is necessary, just as
in the case of conventional spontaneous symmetry breaking.
This generalizes to continuum regularizations with chiral U(1) invariance
such as, for example, momentum-cutoff schemes (see e.g. \rcite{ZJ})
or the dimensional-reduction scheme of \rcite{JJR}.

The main motivation for our lattice-regularization method is that
it may ultimately provide a non-perturbative definition of
(anomaly-free) chiral gauge theories.
These theories are notoriously difficult to regularize
in a gauge-invariant way. In particular,
dimensional regularization is not a gauge-invariant regulator in this case.
A gauge-invariant {\it perturbative} regularization for chiral gauge
theories has been found only recently in the context
of lattice gauge theory~\cite{LPT,SZ}. Beyond perturbation theory,
it is not known if non-abelian chiral gauge theories can
be regularized in a gauge-invariant manner.
(For a review of recent work in this direction see \rcite{ML}.)

According to the gauge-fixing approach, chiral gauge theories
are defined as the continuum limit of a lattice theory whose action contains
a covariant gauge-fixing term and counter terms~\cite{roma,sml,bgs,nona}
(see also \rcite{LAT97} for a pedagogical presentation, and \rcite{mgrev}
for a recent review on lattice chiral gauge theories).
Carefully chosen irrelevant terms in the lattice action
are essential for the existence and the continuity of the phase transition
where the continuum limit is taken.
In the abelian case, we showed that the lattice fermions are
indeed chirally coupled to the gauge field,
and that perturbation theory provides a valid description of the
critical point~\cite{bgs}.

A fully non-perturbative generalization of the gauge-fixing approach
to non-abelian theories is not a simple task, since
one has to confront the issue of Gribov copies.
This important problem will not be addressed here (see \rcite{nona}
for recent progress). Still, the method generates a systematic expansion
around the classical vacuum and, provided the fermion spectrum is
gauge-anomaly free, it provides a consistent
regularization in perturbation theory.
By invoking the familiar machinery of collective coordinates~\cite{thft},
it can be used to generate a systematic expansion
around other classical solutions. In particular,
it is possible to carry out an analytic calculation in an instanton
background. This allows us to address the question,
first raised in~\rcite{TB}, of how fermion-number violating processes
are realized in the gauge-fixing approach.

Let us explain the issue in more detail.
The simplest chiral fermion action used in the gauge-fixing approach,
the so-called chiral Wilson action,
utilizes a set of right-handed {\it spectator}
fields $\c_R^i$, one per each left-handed field $\j_L^i$
(see Appendix~C for the precise definition).
The role of the spectators is to avoid fermion doubling
on the lattice~\cite{W,KS,NN}.
Thanks to a fermion-shift symmetry
they can be proven to decouple in the continuum limit~\cite{GP}.

Each term in the lattice action has one $\bj_L$ or $\bc_R$,
and one $\j_L$ or $\c_R$ field. The chiral Wilson action is therefore
invariant under a common U(1) rotation of {\it all} fermion fields.
The Noether current associated with this symmetry,
\beq
  \jllt_\m \sim \sum_i
  (\bji_L \,\s_\m\, \j_L^i + \bci_R \,\bar\s_\m\, \c_R^i + \cdots) \,,
\lbl{jlatt}
\eeq
is exactly conserved on the lattice (the dots
stand for lattice terms with no continuum counterpart, see Appendix~C).
But since the lattice action is not gauge (nor BRST) invariant
in the gauge-fixing approach, the conservation of $\jllt_\m$
is consistent with our earlier general comments.
Following \rcite{DM}, we have verified through an explicit one-loop
lattice calculation~\cite{pt} that the current $\jllt_\m$ indeed
reduces in the continuum limit to the current $\hjl_\m$ defined in \seeq{hatj}.

The perturbative results of \rcites{DM,pt} are, however, not enough
to resolve the following puzzle, that we will refer to as
the Banks paradox~\cite{TB}. In short, the paradox stems from
the fact that exact U(1) invariance implies exact conservation of
the corresponding U(1) {\it charge}, namely, of fermion number.
To be precise, consider a finite-volume lattice
and assume that the boundary conditions respect the U(1) symmetry.
The Ward identity that corresponds to a global U(1) rotation reads
\beq
  \svev{\d\co} = q \svev{\co} = 0 \,.
\lbl{WI0}
\eeq
Here $q$ is the fermion-number, or the U(1) charge, of $\co$.
We stress that, on a finite lattice, this is a rigorous result.
(The invariance of the lattice measure follows trivially from the fact
that there is an equal number of $d\j_L$ and $d\bj_L$ Grassmann integrals,
as well as an equal number of $d\c_R$ and $d\bc_R$ ones.)

The Ward identity\seneq{WI0} states that
a fermion correlation function $\svev{\co(x_1, x_2,\ldots)}$ can be non-zero
only if $q=0$, namely if the number of $\j_L$ and $\c_R$ fields
is equal to the number of $\bj_L$ and $\bc_R$ fields.
This is true for any finite lattice spacing $a$ and any finite volume,
and therefore also after the continuum limit $a\to 0$
and the infinite-volume limit have been taken.
Moreover, in the continuum limit the numbers of $\c_R$ and $\bc_R$
fields must by themselves be equal,
since the spectator field decouples~\cite{GP}.
Hence the numbers of $\j_L$ and $\bj_L$ fields must be equal too.
We have thus reached the paradoxical conclusion that,
even though the lattice-fermion spectrum is chiral~\cite{bgs},
all fermion-number violating amplitudes vanish!
In other words, 't~Hooft vertices~\cite{thft} do not seem to occur.
If the gauge-fixing approach would be utilized to define a vector-like
theory such as massless QCD,
the same reasoning would seem to lead to the erroneous
conclusion that the U(1) axial charge is conserved in all physical processes.

One can avoid the Banks paradox by adding to the lattice action
a mass term, $m_{ij} \j_L^i \j_L^j + h.c.$, for the physical fermions.
This allows unequal numbers of $\j_L$ and $\bj_L$ fields to
be compensated by insertions of the mass term.
The broken, global U(1) Ward identity in a finite (lattice) volume is now
\beq
  q \svev{\co} = 2 m_{ij} \svev{
  \bigg(\sum_x \j_L^i(x) \j_L^j(x) - h.c.\bigg)\co} \,.
\lbl{WIm}
\eeq
The original lattice-fermion action corresponds to the limit $m \to 0$,
where $m$ denotes generically the magnitude of the mass terms.
We see that fermion-number or axial-charge violating amplitudes can
be non-zero provided they behave like $m/m$ in the limit $m \to 0$.
Now, we would not expect an ``$m/m$'' behavior in a finite volume.
(Being valid for $m=0$, \seeq{WI0} in fact implies that this is impossible
in the presence of a lattice cutoff.) The conclusion is that,
in order to reproduce correctly 't~Hooft vertices,
the infinite-volume limit and the $m\to 0$ limit {\it must not commute}.
Hence, the U(1) lattice symmetry must be broken {\it spontaneously}.

Anticipating spontaneous symmetry breaking (SSB) of the U(1) lattice symmetry
we define the thermodynamical limit as the infinite-volume limit,
followed by the limit $m\to 0$.
In this limit the momentum-space U(1) Ward identity reads,
for any $p_\m \ne 0$,
\beq
  i p_\m \svev{\tjllt_\m(p)\; \co} = q \svev{\co} \,.
\lbl{wi}
\eeq
Here $\tjllt_\m(p)$ is the Fourier transform of $\jllt_\m(x)$.
Unlike \seeq{WIm}, because here $p_\m \ne 0$,
the explicit $m$-dependent term now vanishes for $m\to 0$.
This is explained in more detail in the discussion section.
The identity\seneq{wi} holds
in the regularized theory as well as in the continuum limit.
(On the lattice, the $p_\m$ factor on the left-hand side is modified
by $O(a p^2)$ terms. Again, remember that there is
no room for an anomalous term since
the current $\jllt_\m$ is exactly conserved on the lattice.)

Since 't~Hooft vertices do exist, this means that there are
operators for which the right-hand side of \seeq{wi}
is non-zero. Hence the left-hand side must contain a {\it Goldstone pole}.
As explained earlier, in the continuum limit
the Goldstone pole comes from the $K_\m$ part
of the conserved U(1) current, and is unphysical~\cite{KgSs,SC,thftrev}.

In this paper we calculate instanton-sector fermion correlation functions
on the lattice, in the semi-classical approximation.
We start from a lattice-fermion action with an additional, small,
mass term that breaks explicitly the unphysical U(1) symmetry.
Taking the infinite-volume and continuum limits, followed
by the $m\to 0$ limit, we show that
the anticipated 't~Hooft vertices are recovered.
The ``$m/m$'' nature of fermion-number and axial-charge violating
amplitudes is manifest in our calculation.

The paper is organized as follows.
In order to minimize technicalities we begin in Sect.~2 with
one-flavor QCD where, instead of the usual gauge-invariant lattice definition,
we define the theory via the gauge-fixing approach in the special case that the
(left-handed) fermion spectrum happen to contain one field in
the fundamental representation and one field in the complex conjugate one.
In Sect.~3 we work out the anomaly-free SO(10) theory
as the prototype of a truly chiral gauge theory.
Our conclusions are summarized and discussed in Sect.~4.
In particular, we show that the phase of the 't~Hooft vertex follows
the phase of the applied mass term, thus demonstrating explicitly
the existence of the continuously degenerate ground states
associated with SSB.
Notations are listed in Appendix~A,
elements of SO(10) group theory are discussed in Appendix~B, and
lattice definitions are collected in Appendix~C.
The construction of propagators in the presence of
approximate zero modes is discussed in Appendix~D.

\vspace{5ex}
\noindent {\large\bf 2.~One-flavor QCD using the gauge-fixing approach}
\secteq{2}
\vspace{3ex}

We begin with the simple example of one-flavor massless QCD,
an SU(N) gauge theory coupled to one Dirac fermion in the
fundamental representation. The anomalous current of \seeq{dj} is
in this case the axial current.
Let us recall what the 't~Hooft interaction
of this theory is. In a fixed instanton background,
the massless continuum Dirac operator $\Sl{D}$ has one left-handed zero mode
$u(x)=P_L u(x)$. Therefore the Weyl fields $\j_L$ and $\bj_R$
each have a zero mode. (In an anti-instanton background,
the Weyl fields with zero modes are $\j_R$ and $\bj_L$.)
The basic axial-symmetry violating correlation function is
\beq
  \svev{\j_L(x)\,\bj_R(y)} = u(x)\, u^\dagger(y) \, \Det' \,,
\lbl{RL}
\eeq
where the expectation value denotes Grassmann integration only,
and ${\rm Det'}$ is the (renormalized)
fermion determinant with the zero mode removed.
Our objective will be to recover this result starting from
a lattice action with exact axial U(1) invariance.

The remaining step in a complete semi-classical calculation
is the integration over the gauge and ghost fields.
This raises no new conceptual issues and therefore we will skip
the details. We recall that the integration
(or lattice-sum) over the instanton position
recovers momentum conservation.
The gaussian integration over the non-zero gauge, fermion and ghost
fluctuations leads to the replacement of the lattice's bare instanton action,
$8\p^2/g_0^2$, by the renormalized one, $8\p^2/g_r^2(\r)$,
where $\r$ is the instanton's size.

One-flavor QCD has a gauge-invariant lattice definition.
When using ordinary Wilson fermions, the lattice fermion action
is not invariant under axial transformations, and the paradox described
in the introduction does not arise. (In the continuum limit
one reproduces the axial anomaly~\cite{KS}, while non-singlet
axial symmetries are recovered~\cite{Betal}.)
In principle, it should be possible to define one-flavor QCD
using the gauge-fixing approach, too. While this has
many disadvantages compared to the gauge-invariant definition,
it has the interesting property that the paradox described
in the introduction {\it occurs}. By working out one-flavor QCD
we are able to address, with minimal technicalities,
the main issue of this paper --- how a global
symmetry of the lattice path integral can be broken in the continuum limit.

The lattice construction of the chiral Wilson action begins with enumerating
the left-handed fields of the target theory. For one-flavor QCD
we have two Weyl fields $\j_L$ and $\j_L^c$ in the fundamental
and anti-fundamental representations respectively.
As explained in the introduction, one also needs two right-handed
spectator fields that decouple in the continuum limit.
These may be denoted $\c_R$ and $\c_R^c$.
In the case at hand, we may take advantage of the Dirac nature
of the target theory, and trade the left-handed anti-fundamental field
with a right-handed fundamental one
$\j_L^c \to \bj_R$, $\bj_L^c \to \j_R$.
With a similar tradeoff for the corresponding spectator field,
the lattice fermion action density can be written in the following matrix form
\beq
  \raisebox{0ex}{$
  \left(
  \begin{array}{cccc}
    \bc_R        & \bj_L     & \bj_R        & \bc_L
  \end{array}
  \right)
  $}
  \left(
  \begin{array}{cccc}
    0           & 0                & -{a\over 2}\hb & \bar\s_\m\hp_\m \\
    0           & m                & \s_\m\hd_\m    & -{a\over 2}\hb  \\
 -{a\over 2}\hb & \bar\s_\m \hd_\m & m              & 0               \\
    \s_\m\hp_\m & -{a\over 2}\hb   & 0              & 0
  \end{array}
  \right)
  \left(
  \begin{array}{c}
    \c_L           \\ \j_R        \\ \j_L          \\ \c_R
  \end{array}
  \right)\,.
\lbl{cW}
\eeq
We use hats to denote lattice derivatives. (For the precise definitions
of lattice derivatives and currents see Appendix~C.)
Observe that the middle two-by-two block in the above matrix operator
resembles the massive continuum Dirac operator.

For orientation, we recall that in the conventional definition
of Wilson fermions there are of course no spectator fields, and covariant
Wilson terms are placed in the same block-entries as the mass terms
in the above expression. This removes the doublers in a gauge invariant way,
while axial symmetry is lost.

In order to later accommodate truly chiral gauge theories,
the doublers are removed here
by introducing spectator fields and coupling them to the original
fermions via the (free) lattice laplacian $\hb$.
Since the Wilson terms now couple fields with different
gauge-transformation properties,
they lead to breakdown of gauge invariance~\cite{ssm}.
In lattice perturbation theory
gauge invariance is regained by adding suitable counter terms,
and the renormalized diagrams describe one interacting Dirac field, the quark,
and one free Dirac field, the spectator.
As is usually the case for symmetries broken by the lattice regularization,
the above is true provided the external momenta are vanishingly small
in lattice units. The choice of a free lattice laplacian in \seeq{cW}
implies the shift symmetry of the spectator field~\cite{GP},
which reduces considerably the number of counter terms.
In particular, there are no counter terms of the form
$\bj_R\,\chi_L$, etc., and the spectator field decouples
in the continuum limit.

Let us now examine the U(1) symmetries of the action\seneq{cW}.
Dropping both the Wilson and the mass terms,
the action would be invariant under four separate U(1)'s ---
a fermion-number symmetry for each Weyl field.
With the Wilson terms in place, the action is still invariant under two U(1)'s.
Finally, for $m \ne 0$, only the invariance under a common U(1)
rotation is left. This invariance corresponds to
the baryon-number symmetry of QCD.

The additional U(1) symmetry at $m=0$ transforms $\c_L$
and $\j_R$ with (say) charge $+1$, and $\c_R$ and $\j_L$
with charge $-1$. This is the chiral symmetry that leads to the Banks
paradox~\cite{TB}. The existence of this lattice symmetry
would seem to lead to the (erroneous) conclusion that the axial charge
is conserved in massless (one-flavor) QCD.
As mentioned in the introduction, \rcites{DM,pt}
already showed that the anomaly appears in the triangle diagram
as expected. But this still does not explain how the axial charge
is {\it not} conserved in physical processes.

We will now answer this question through an explicit calculation.
We calculate the lattice-fermion two point function in the semi-classical
approximation for (small) $m>0$. Taking the infinite-volume and
continuum limits, and finally sending $m\to 0$,
we find that the 't~Hooft interaction\seneq{RL} is recovered.

As discussed in the literature~\cite{KgSs,SC,thftrev},
the continuous degeneracy of ground states associated with SSB
of the axial U(1) is parametrized by the vacuum $\th$-angle.
The chiral Wilson action\seneq{cwl} is invariant under a CP transformation,
and this remains true in the presence of the mass term introduced in
\seeq{cW} above. Therefore the calculation in this section
(as well as in section~3) corresponds to a vacuum
angle $\th=0$. The case of general $\th$-angle is explained
in the discussion section.

We start with a continuum {\it regular-gauge} instanton field $A_\m(x)$
whose size $\r$ is very large in lattice units, $\r\gg a$.
(Singular-gauge instantons are suppressed in the gauge-fixing approach
by the irrelevant terms in the lattice action, see Appendix~C.)
The lattice gauge field may be defined as $U_\m(x)=\exp(iaA_\m(x))$.
This is a  smooth configuration. (By this we mean that $U_\m(x)-I = O(a/\r)$
and $U_\m(x)-U_\m(x+\hat\n) = O((a/\r)^2)$.)
For this lattice gauge field and fixed $m>0$,
we denote the matrix operator in \seeq{cW} by $\g_5\cm$.
Note that according to this definition $\cm$ is hermitian.

In the formal continuum limit, the lattice operator in \seeq{cW}
goes over to a continuum Dirac operator $D(m)=\g_5 H(m)$
which depends on the original instanton field $A_\m(x)$.
One obtains $D(m)$ by dropping
the (irrelevant) Wilson terms and replacing the lattice difference
operators $\hat\partial_\m$ and $\hat{D}_\m$ by the corresponding
continuum derivatives. For $m>0$, $D(m)$ describes a massive quark
(made of $\j_{R,L}$) whose Dirac operator is $\Sl{D}+m$,
and a decoupled, free massless spectator field (made of $\c_{R,L}$).
Thus, $D(m)$ has no zero modes.
We will denote by $G(m)$ the propagator of the hermitian operator $H(m)$.
Both $\cm^{-1}$ and $G(m)$ admit a standard spectral decomposition.

Let us now be more precise about how the physical matrix element
is obtained. One has to multiply the correlation function in \seeq{RL}
by (normalized) wave-functions $f_1^\dagger(x)$ and $f_2(y)$
and integrate (or sum) over $x$ and $y$.
The physical observable is the gauge-field functional average
of $\Det(\cm)\sbra{f_1}\cm^{-1}\g_5\sket{f_2}$.
We will denote the generic virtuality of the external legs by $Q^2$.
As explained below, the matrix element is dominated by
instantons of size $\r^2 \sim Q^{-2} \gg a^2$, where the last inequality
follows because in the continuum limit  $a^2 Q^2 \to 0$.

Since the fermion determinant will be $O(m)$, we are interested only
in the $O(1/m)$ piece of the propagator(s). We claim that
\beq
  \lim_{a\to 0}\,
  \sbra{f_1} \cm^{-1} \g_5\sket{f_2}^{\rm singular}_{\rm lattice}
  = \sbra{f_1} G(m) \g_5\sket{f_2}^{\rm singular}_{\rm continuum}\,.
\lbl{MGm}
\eeq
Here ``singular'' denotes the $O(1/m)$ piece.
The reason why \seeq{MGm} is true is that a non-zero $m$ affects
only eigenfunctions with eigenvalues $\l$ in the region $|\l|\leqx m$.
Indeed, for all (lattice or continuum) eigenfunctions with, say,
$\l^2 \ge Q^2$, the effect of $m>0$ will be bounded by $m^2/Q^2$
to some positive power.
Therefore they do not contribute to the $O(1/m)$ term.
For $\l^2 \le Q^2$, the difference between each continuum eigenfunction
and the corresponding lattice eigenfunction
is bounded by $a^2 Q^2$ to some positive power.
Since $m>0$, the inverse eigenvalues $\l^{-1}$ are bounded from above,
and the contribution of the entire low-energy lattice spectrum approaches
smoothly the continuum one for $a\to 0$.
(For the pairing of the lattice and the continuum eigenfunctions
we may momentarily assume a very large, but finite, volume,
thus making the spectrum discrete; alternatively
\seeq{MGm} can also be justified directly in the infinite-volume limit.)

In the infinite-volume limit, the massive continuum propagator satisfies
\beq
  G(x,y;m) \g_5 =  {1\over m}\, u(x)u^\dagger(y) + O(1) \,.
\lbl{msv}
\eeq
Here we show explicitly
only the term that diverges for $m\to 0$ (compare \seeq{Gx0}).
With the understanding that the matrix element is to be taken
between smooth wave-functions $f_{1,2}$ as described above, we thus have
\beq
  \lim_{a\to 0} \cm^{-1}(x,y)\g_5
  =  {1\over m}\,u(x)\,u^\dagger(y) + O(1) \,.
\lbl{m}
\eeq
For the fermionic determinant similar arguments lead,
after renormalization, to
\beq
  \lim_{a\to 0}\, \Det(\cm) = m (\Det' + O(m)) \,.
\eeq
The explicit factor of $m$ again comes from the (approximate) zero mode,
while the $O(m)$ terms account for the change in the continuous spectrum
due to $m$. Putting this together we thus obtain
\beq
  \svev{\j_L(x)\,\bj_R(y)}
  = m \left(\Det' + O(m) \right)
    \left( {1\over m}\, u(x)\, u^\dagger(y) + O(1)\right)\,.
\lbl{RLa}
\eeq
Finally, taking the limit $m\to 0$ we recover \seeq{RL}.
\seEq{RLa} reveals the ``$m/m$'' nature of the 't~Hooft interaction.

The familiar \seeq{msv} above is particularly simple, and has been invoked
primarily for pedagogical reasons.
In Appendix~D we show how to handle perturbations that lift the zero modes,
but are not proportional to the identity matrix and/or
are not spatially constant. This more general formalism
will be necessary in the next section.
For a few more details on the calculation
of the determinant see {\it Comment~2} in Appendix~D.

Our instanton calculation was done in the semi-classical approximation,
as is routine in the continuum. Since we have somewhat expanded its scope
by using a specific lattice-regularization method,
we will briefly review the justification for the semi-classical approximation.

Consider a fermion-number violating amplitude
with only a minimal number of fermions,
and no other particles, as the external legs.
(Here we wish to avoid the controversy about whether the fermion-number
violating cross section could become large at very high energies
due to multi-boson final states.)
As before, denote the generic virtualities of the external legs by $Q^2$.
For instanton size $\r^2 \gg Q^{-2}$,
the overlap of the zero modes with the wave-functions on the external legs
will provide a strong damping factor. Hence the saddle point $\r_{\rm sp}$
of the integration over the instanton's size is $\r_{\rm sp}^2 \sim Q^{-2}$.
If $Q^2$ is much larger than the confinement scale,
the running coupling $g_r=g_r(\r_{\rm sp} \sim \sqrt{Q^{-2}}\,)$ is small.
This justifies the use of the one-instanton approximation.

Ultimately, the most visible consequence of the anomaly in one-flavor
QCD is that the lightest pseudo-scalar state (the ``$\h'$ meson'')
is {\it not} light compared to the confinement scale
(see \eg \rcite{thftrev}).
The chiral-symmetry breaking effect obtained from the
semi-classical instanton calculation
is much smaller since it is controlled by the small parameter
$\exp(-8\p^2/g_r^2(\r_{\rm sp}))$. We resort to this
deep euclidean regime, because only there are we able to apply analytic methods
to accurately calculate the consequences of the anomaly.

The above considerations have to do with the asymptotically-free nature
of the Yang-Mills coupling, and therefore they are completely independent
of the regularization method. Moreover, our explicit calculation
has demonstrated that no uncontrolled lattice artifacts occur.
Finally, we note that the discretization of regular-gauge instantons does yield
gauge-field configurations that fail to satisfy the lattice Yang-Mills
equation of motion, but only by a small amount $O(a/\r_{\rm sp})$.
Instanton-sector Feynman rules that generate a systematic expansion
in $g_r^2(\r_{\rm sp})$ can be derived
in the presence of an approximate classical solution, see \eg \rcite{susy}.

\vspace{5ex}
\noindent {\large\bf 3.~Chiral gauge theories}
\secteq{3}
\vspace{3ex}

The lesson of the previous section is that
an 't~Hooft vertex can be interpreted as an order parameter for
the spontaneous breaking of the U(1) chiral symmetry
in a regularization scheme where chiral (but not gauge) invariance
is preserved. The introduction of a small mass term, which is sent
to zero after the infinite-volume limit was taken, provides the
necessary coupling to an ``external magnetic field'' and
allows the expectation value of an 't~Hooft vertex to be non-zero.
This reasoning is valid both in the continuum
and on the lattice, if one uses the gauge-fixing approach.

The generalization of the previous calculation to 't~Hooft vertices
that violate the fermion-number symmetry of a chiral gauge theory
is relatively straightforward.
Starting from the lattice theory, a mass perturbation
that lifts the fermionic zero modes will again
allow us to keep the (approximate) zero modes under control
while taking the infinite-volume and continuum limits.
Performing next the limit $m\to 0$, we will
recover the 't~Hooft vertices as before.
The only step which may not be obvious, is that
a mass perturbation that lifts all zero modes exists in the continuum.

In this section we demonstrate the existence of the necessary
mass perturbation by working out the example of an SO(10) chiral gauge theory.
(Attempting to construct the necessary mass perturbation
for the most general anomaly-free
chiral gauge theory may be tedious, and the SO(10) example is
general enough to encompass the Standard Model as well as
the most popular Grand Unification schemes.)
In a one-generation SO(10) theory the Weyl fermions reside in
the complex {\sixteen} representation.
We introduce covariant derivatives (${\sstyle M,N}=1,\ldots,10$)
\bqry
  D_\m       & = & \partial_\m + i A_\m^{\ssstyle MN} \S_{\ssstyle MN} \,, \NON
  \bar{D}_\m & = & \partial_\m + i A_\m^{\ssstyle MN} \bar\S_{\ssstyle MN} \,,
\lbl{covder}
\eqry
in the {\sixteen} and the {\sixbar} representations respectively.
The SO(10) generators are defined via
\beq
  {i\over 2}\, [\G_{\ssstyle M}\,, \G_{\ssstyle N} ]
  =
  \half (1 + \G_{11}) \S_{\ssstyle MN}
  + \half (1 - \G_{11}) \bar\S_{\ssstyle MN}\,.
\lbl{smn}
\eeq
We use the 32 by 32 representation of the ten-dimensional gamma matrices
given in Appendix~B. The (continuum) lagrangian is
\beq
  \cl = \bj_L\, \s_\m D_\m \, \j_L \,.
\lbl{lag}
\eeq

In an instanton background there are four left-handed zero modes,
one for each quark or lepton. We will show that a suitable
mass term lifts all four zero modes.
To prepare for the introduction of the mass term
we first rewrite the lagrangian in terms of Majorana-like fermions
\beq
  \J =
  \left(
  \begin{array}{c}
    \vspace{1ex}
    \e\, \cc \, \bj_L^T \\
    \j_L
  \end{array}
  \right) \,, \qquad
  \bJ =
  \left(
  \begin{array}{cc}
    \bj_L  &  \j_L^T \e \, \cc
  \end{array}
  \right) \,,
\lbl{maj}
\eeq
satisfying
\beq
  \bJ \equiv \J^T \, C_4 \, \cc \,,
\lbl{majcond}
\eeq
where $\e$ is the anti-symmetric two-dimensional tensor and $C_4$
is the four-dimensional charge conjugation matrix (see Appendix~A).
The 16 by 16 matrix $\cc$, which is related to the ten-dimensional
charge-conjugation matrix, is defined in \seeq{cc}. It satisfies
$\cc^*=\cc^T=\cc^{-1}=\cc$.
In terms of the Majorana-like fields the lagrangian is rewritten as
\beq
  \cl = {1\over 2} \, \bJ\, D_0\, \J \,,
\lbl{majlag}
\eeq
where
\beq
  D_0 = \g_\m (D_\m P_L + \bar{D}_\m P_R )
  =
  \left(
  \begin{array}{cc}
    0                   & \s_\m D_\m \\
    \bar\s_\m\bar{D}_\m & 0
  \end{array}
  \right)\,.
\lbl{majD}
\eeq
Note that $D_0^\dagger \ne -D_0$:
unlike the QCD case, $D_0$ is not anti-hermitian.
One can show that
\beq
  D_0^\dagger C_4 \cc = C_4 \cc D_0^* \,.
\lbl{majrel}
\eeq
Appendices~A and~B contain a number of useful relations
which have been used above.

Equipped with the Majorana formulation
we introduce a mass term
\beq
  {m\over 2}\;\bJ\J
  = {m\over 2} \left(\j_L^T \e\,\cc \j_L + \bj_L \e\,\cc \bj_L^T\right)\,.
\lbl{mass}
\eeq
The mass term breaks explicitly the fermion-number symmetry
and, in the limit $m \to 0$,
provides the ``seed'' for spontaneous symmetry breaking.
(The mass term also breaks the chiral gauge invariance, see below.)
The fermion operator becomes
\beq
  D(m) = D_0 + m \,.
\lbl{Dm}
\eeq
\seEq{majrel} holds for $D(m)$ too.

We will soon prove that $D(m)$ has no zero modes, for $m\ne 0$. But first,
we give a simple physical explanation why this should be expected.
Observe that
\bqry
  \cc \otimes I = -i \, C_{10} \, \G_{10} \,,
\nonumber
\eqry
where $I$ is the two-by-two identity matrix.
Introducing a 32-component spinor $\J'$ whose first 16 components are
equal to $\J$ we may write
\beq
  \bJ\J \equiv \J^T C_4 \cc \, \J
  = -{i\over 2}\, (\J')^T C_4 C_{10} \, \G_{10} (1 + \G_{11}) \J' \,.
\eeq
Because of $\G_{10}$,
the mass term can be thought of as coming from the vacuum expectation value
of a Higgs field in the \tenrep representation.
This vacuum expectation value breaks SO(10) down to SO(9).
Since all spinor representations of SO(9) are real,
the fermions can acquire Majorana masses consistently with SO(9) invariance.
Moreover, the sixteen-dimensional representation of SO(9) is irreducible,
and therefore {\em all} sixteen fermions acquire a Majorana mass.

We will now show in more detail
that there are no exact zero modes for $m\ne 0$,
and that the fermion-number-violating 't~Hooft interaction
is recovered in the limit $m\to 0$. We describe the main steps here,
relegating further technical details to Appendix~B.
In order to obtain information
on the fermion propagator for $m\ne 0$
we will need the general formalism of Appendix~D,
which applies to hermitian operators. We will thus consider
the following hermitian operator (and corresponding propagator)
\beq
  \cd(m)
  =
  \left(
  \begin{array}{cc}
    0            & D(m)  \\
    D^\dagger(m) & 0
  \end{array}
  \right)
\,,\qquad
  \cg(m)
  =
  \left(
  \begin{array}{cc}
    0    & G^\dagger(m)  \\
    G(m) & 0
  \end{array}
  \right) \,.
\lbl{hrmt}
\eeq
Note that $D(m)$ carries a four-component spinor index,
and $\cd(m)$ carries an eight-component spinor index.
Let us first enumerate the zero modes for $m=0$.
There are the four original left-handed zero modes $u_i$
that belong to the {\sixteen} and satisfy $\s_\m D_\m\, u_i = 0$.
In addition, define
\beq
  v_i = - \e \, \cc \, u_i^* \,.
\lbl{uv}
\eeq
Using the left-handedness of $u_i$ and \seeq{majrel} one has
\beq
  D_0^\dagger
      \left(\begin{array}{c}
         0  \\
         v_i
      \end{array}\right)
  = - C_4 \cc \left(D_0
      \left(\begin{array}{c}
         0  \\
         u_i
      \end{array}\right)
    \right)^* = 0\,.
\lbl{ustar}
\eeq
Therefore the $v_i$ are left-handed zero modes of $D_0^\dagger$
that belong to \sixbar. The propagator $\cg_0(x,y)$ is orthogonal
to all eight zero modes and satisfies
\beq
  \cd_0\,\cg_0(x,y) = \d^4(x-y) - \cp(x,y) \,,
\lbl{clsr}
\eeq
where the zero-mode projector is
\beq
  \cp(x,y)
  =
  \left(
  \begin{array}{cccc}
    0 &                      0  &  0  &  0  \\
    0 &  v_i(x) v_i^\dagger(y)  &  0  &  0  \\
    0 &                      0  &  0  &  0  \\
    0 &                      0  &  0  &  u_i(x) u_i^\dagger(y)
  \end{array}
  \right) \,.
\lbl{proj}
\eeq
{}From these equations one can read off the relations satisfied by
the chiral propagator:
\bqry
  D_0\, G_0(x,y)      & = & \d^4(x-y) - P_L \, v_i(x) v_i^\dagger(y) \, P_L \,,
\NON
  G_0(x,y) \lvec{D}_0 & = & \d^4(x-y) - P_L \, u_i(x) u_i^\dagger(y) \, P_L \,.
\nonumber
\eqry
The derivative acting to the left has a minus sign.

We now turn to $m \ne 0$. As explained earlier, the fermion-number symmetry
is broken explicitly. This is reflected by the fact
that $G(m)$ does not anti-commute with $\g_5$
(compare \seeq{contract} below).
An inspection of \seeqs{Dm} and\seneq{hrmt}
reveals that to first order, the mass perturbation
can have non-zero matrix elements
only between a {\sixteen} and a {\sixbar} zero-mode. Let
\beq
  \l_{ij} = \sbra{v_i} m \sket{u_j} = m \sbraket{v_i}{u_j}\,.
\lbl{luv}
\eeq
We find, using \seeq{Gx0},
\beq
  G(m) = \sket{u_i} \l^{-1}_{ij} \sbra{v_j} + O(1) \,.
\lbl{Olmbd}
\eeq
The first term on the right-hand side is $O(1/m)$.
Furthermore, using \seeq{uv},
\beq
  \l_{ij} = m \int u_i^T  \e \, \cc  u_j \,,
\lbl{lt}
\eeq
which implies that $\l_{ij}$ is antisymmetric.
In the zero-mode sector,
the Majorana-fermion determinant is the analytic square root (pfaffian)
\beq
  {\rm det}^{1/2}(\l) = {1\over 8} \e_{ijkl} \, \l_{ij}  \, \l_{kl} \,.
\lbl{Pf}
\eeq
In Appendix~B we prove that $\det(\l) \ne 0$ for (almost) every
embedding of the instanton in SO(10).

Next consider correlation functions in the instanton sector.
In terms of the original Weyl fields,
\beq
  G(m)
  =\vev{\Psi\overline\Psi}=
  \left(
  \begin{array}{cc}
    \vev{\e\, \cc  \bj_L^T\,\bj_L} &
       \vev{\e\, \cc  \bj_L^T\,\j_L^T \e\, \cc} \vspace{1ex}  \\
    \vev{\j_L \,\bj_L}     &  \vev{\j_L \,\j_L^T \e\, \cc}
  \end{array}
  \right) \,.
\lbl{contract}
\eeq
Let us first see what happens if we saturate two fermion fields
$\j_L(x)\j_L(y)$ by the $O(1/m)$ part
of the propagator. Using \seeqs{uv}, \seneq{Olmbd},
and noting the lower-right entry of \seeq{contract}, we obtain a factor
\beq
  u_i(x) u_j(y) \l^{-1}_{ij} \,.
\eeq
By itself, this will give a vanishing result in the limit
$m \to 0$ because ${\rm det}^{1/2}(\l)$ is $O(m^2)$.
Next consider saturating the product of four fields
$\j_L(x)\j_L(y)\j_L(z)\j_L(w)$. Summing over all possible contractions
and paying attention to Fermi statistics we get
\beq
  u_i(x) u_j(y)  u_k(z) u_l(w)
  \left( \l^{-1}_{ij}\l^{-1}_{kl} + \l^{-1}_{ik}\l^{-1}_{lj}
  + \l^{-1}_{il}\l^{-1}_{jk} \right) \,.
\eeq
The expression in parentheses is completely anti-symmetric
in the four indices $i,j,k,l$.
(If we would try to saturate $2n$ fermion fields
for some $n>2$ with the $O(1/m)$ part of the propagator,
the result would be identically zero due to the anti-symmetrization.)
To evaluate the sum we write
\beq
  \left( \l^{-1}_{ij}\l^{-1}_{kl} + \l^{-1}_{ik}\l^{-1}_{lj}
  + \l^{-1}_{il}\l^{-1}_{jk} \right)
  = c \, \e_{ijkl} \,,
\eeq
and contracting with another $\e_{ijkl}$ we find
\beq
  c = {1\over 8}\, \e_{ijkl} \l^{-1}_{ij}\l^{-1}_{kl}
  = {\rm det}^{1/2}(\l^{-1})\,.
\lbl{vrtx}
\eeq
This cancels against \seeq{Pf}, and,
in the limit $m \to 0$, one is left with
\beq
  \svev{\j_L(x)\j_L(y)\j_L(z)\j_L(w)}
  =
  \e_{ijkl}\, u_i(x) u_j(y)  u_k(z) u_l(w)\, \Det' \,,
\eeq
which is the expected 't~Hooft vertex.

The mass term\seneq{mass} breaks not only
the unwanted fermion-number symmetry; it also breaks chiral gauge invariance.
This, however, does not lead to any disaster; in fact, we know that in
the UV-regulated theory the gauge symmetry is already broken by the
regulator.
The crucial point is that, in a covariantly gauge-fixed theory,
the ultra-violet behavior of the vector-boson propagator is $1/p^2$
for all polarizations. Consequently, the theory
remains renormalizable even if terms that break gauge
invariance are added.

Moreover, the addition of a mass term
does not change the nature of the coupling of the theory.
The one-loop beta function is unaffected
by the mass term, and so the theory is still asymptotically free.
Also, provided we are careful to employ a mass-independent
renormalization prescription, universality of the
renormalized coupling should be preserved.
The same considerations imply that the continuum limit of the lattice theory
should exist for $m \ne 0$, too. In the renormalized theory,
a fermion-mass term is expected to induce a vector-boson
mass term, and so unitarity will be violated by $O(m)$ effects.
In the limit $m \to 0$ (keeping physical scales fixed)
unitarity should be recovered.

\vspace{5ex}
\noindent {\large\bf 4.~Discussion}
\secteq{4}
\vspace{3ex}

The classically-conserved chiral U(1) symmetry
is not preserved by any gauge-invariant regularization,
and in the quantized theory physical observables exist that violate
this symmetry.
In this paper we have considered an important aspect
of regularization methods which are not gauge invariant but,
instead, respect chiral U(1) invariance. Our concrete motivation to do so
is the gauge-fixing approach to (chiral) lattice gauge theories.
We showed that even with a chiral-U(1) invariant regulator,
a careful treatment of the {\it infra-red} limit reproduces correctly
the gauge invariant, chiral-symmetry-violating 't~Hooft vertices.
In essence, our conclusions are as follows.

{\it (I) Since the chiral U(1) symmetry is preserved by the regularization
but is not respected by physical amplitudes, it must be broken spontaneously.
(II) Therefore, in order to obtain the physical amplitudes,
one should introduce a mass perturbation that breaks
the chiral symmetry explicitly, and take the limit $m\to 0$ after
the infinite-volume limit. (III) The Hilbert space of the gauge-fixed theory
will contain a corresponding Goldstone pole, but it is unphysical, because
it originates from an unphysical (gauge non-invariant) conserved current.}

The second and third statements are
closely tied to the first one. When there is SSB
the thermodynamical limit is defined
by introducing an ``external magnetic field'' (here the mass term)
which is switched off after the infinite-volume limit has been taken.
As for the existence of a massless pole,
it is a consequence of locality and the Goldstone theorem which,
in the present context, has been noted before in \rcites{KgSs,SC}.

The mass perturbation allows us to avoid the Banks paradox,
namely the apparent conflict between the symmetries of the regularized
theory and of the physical amplitudes.
Starting from a gauge-fixed lattice theory
that has the unwanted chiral symmetry, we have demonstrated through explicit
examples how this mechanism works for the anomalous axial
symmetry of QCD-like theories, and for fermion-number-violating
processes in chiral gauge theories.

Zero modes belong to the low end of the fermion spectrum which,
in general, is sensitive to infra-red details such as having $m\ne 0$,
finite versus infinite volume, and choices of boundary conditions.
Keeping $m>0$ provides an infra-red regularization for all
fermionic correlation functions, and allows us to take
the infinite-volume limit without difficulty.
Once in infinite volume, we obtain the correct 't~Hooft vertices
because we evidently have the correct number
of (approximate) zero modes.

As explained earlier, because of CP invariance, the vacuum angle $\th$
was equal to zero in the previous sections.
Considering the one-flavor QCD example for simplicity,
let us examine a mass term with a general axial-U(1) phase.
(The generalization to chiral gauge theories is straightforward;
see \rcite{SeiSt} for a discussion of $\th$-vacua in the context
of the standard Wilson action.)
The lattice action is now defined by replacing the mass term
in \seeq{cW} with a new mass term pointing
in an arbitrary axial-U(1) direction
(the parameter $m$ is real)
\beq
  m\,\bj\, e^{i\th\gamma_5}\, \j = m\left(
    e^{i\th}\, \bj_L \j_R + e^{-i\th}\, \bj_R \j_L
  \right) \,.
\lbl{mth}
\eeq
By applying an axial-U(1) transformation
($\j \to e^{-i\th\gamma_5/2}\, \j$, $\c \to e^{i\th\gamma_5/2}\, \c$, etc.),
and using the invariance of the lattice theory for $m=0$,
we can relate the value of any $\th \ne 0$ correlation function
to its value at $\th=0$.
For the correlation function of \seeq{RL} the result is
\beq
  \svev{\j_L\, \bj_R}_\th = e^{i\th} \svev{\j_L\, \bj_R}_{\th=0} \,.
\lbl{key}
\eeq
The subscript $\th$ refers to the angle of the mass term\seneq{mth}.
(Note that, in \seeq{key}, $\exp(i\th)$ may be re-expressed as $\exp(-iq\th/2)$
where $q=-2$ is the axial charge of $\j_L \bj_R$.)

\seEq{key} is a rigorous result in the lattice theory.
A similar relation holds in the thermodynamical limit.
Hence the order parameter for axial-U(1) symmetry breaking
--- the 't~Hooft vertex ---
acquires a phase equal to the phase of the applied mass term.
This proves that the SSB ground states are indeed
parametrized by the (vacuum) $\th$-angle.

The relevance of $\th$-vacua to the U(1) problem was discussed
in detail in the literature. In a continuum treatment
the path integral can only be expanded perturbatively around
selected classical fields, and
$\th$-vacua must be incorporated ``by hand.''
Here, we showed that the $\th$-angle of the 't~Hooft vertex arises as an
unavoidable consequence of the lattice regularization.

While our explicit lattice calculation does not depend on this,
let us recall some related observations from \rcites{KgSs,SC,thftrev}.
Any sector with a fixed topological charge, such as the one-instanton sector,
is a superposition of all the $\th$-vacua.
But for any $m\ne 0$, however small,
the vacuum energy density is a non-trivial function of $\th$.
This explains why a unique $\th$-vacuum is selected
in the thermodynamical limit.

Had the infinite-volume limit been taken while keeping $m=0$,
the $\th$-vacua would have remained exactly degenerate.
This prescription would yield a vanishing result for
chiral-symmetry violating amplitudes.
But since {\it clustering} would be violated,
this prescription is inconsistent.
Returning to the lattice regularization, this may be explained as follows.
If the lattice volume is finite, $V<\infty$, the limit $m\to 0$ must be
smooth and, hence, independent of $\th$. Namely,
\beq
  \svev{\j_L\, \bj_R}_{m\to 0,V<\infty, \th}
  = \svev{\j_L\, \bj_R}_{m\to 0,V<\infty, \th=0} \,.
\lbl{fV}
\eeq
Since this must be true for all $\th$ simultaneously with \seeq{key},
the conclusion is that the 't~Hooft vertex
vanishes on a finite lattice, if we set $m=0$.
(This argument is really an alternative explanation of the
Ward identity\seneq{WI0}, but makes the role
of the axial phase more explicit.)
An implication is that, clearly, one would have to keep $m\ne 0$
in a numerical simulation in order to recover 't~Hooft vertices.

The 't~Hooft vertices are characterized by an ``$m/m$'' behavior
in the thermodynamical limit.
Given a chiral U(1) Ward identity,
should an ``$m/m$'' behavior be expected from
any other term except the symmetry-breaking expectation value?
The answer is no.
As a concrete example consider the following momentum-space
Ward identity in one-flavor QCD with $m\ne 0$,
as defined on the lattice via the gauge-fixing approach (Sect.~2).
For $|pa| \ll 1$ it reads (compare \seeq{parcons})
\beq
  ip_\m \svev{\tjlt_{5\m}(p)\, J_5}
  = -2m \svev{\tilde{J}_5(p)\; J_5} + 2\svev{\bj\j} \,.
\lbl{WIpm}
\eeq
In this equation, $\jlt_{5\m}$ is the conserved U(1) axial current in the
limit $m\to 0$, and $\bj\j$ and $J_5$ are the local scalar and pseudo-scalar
lattice densities. As in \seeq{wi} the tilde denotes Fourier transform.
The expectation value of $\bj\j$ corresponds to the
limit $x=y$ in \seeq{RL}. While in this limit the semi-classical
calculation ceases to be reliable, we take the non-zero result for the
't~Hooft vertex as an evidence
that $\svev{\bj\j} \ne 0$.

The contribution of an approximate Goldstone pole to
$\svev{\tilde{J}_5(p)\; J_5}$
should be proportional to $(p^2 + vm)^{-1}$
where $v$ is a dimensionful constant. The corresponding contribution
to the Ward identity goes like $m/(p^2 + vm)$
and vanishes in the limit  $m\to 0$.
The contribution of all other excitations to $\svev{\tilde{J}_5(p)\; J_5}$
should be less infra-red singular. Therefore nothing that behaves like
``$m/m$'' should arise from $\svev{\tilde{J}_5(p)\; J_5}$, so long
as we are careful to keep the momentum {\it not} strictly zero.
Indeed, sending $p\to 0$ as a {\it limit} is an inherent part
of the Goldstone theorem (see for example \rcite{GHK}).
For $m\to 0$ we thus obtain
\beq
  ip_\m \svev{\jlt_{5\m}(p)\, J_5}
  = 2\svev{\bj\j} \,.
\lbl{WIp}
\eeq
This equation is a special case of the Ward identity\seneq{wi}.

The problem of fermion-number violation in lattice chiral gauge
theories was previously also addressed in \rcite{BHS}.
In the (axial) Schwinger model~\cite{JS}, these authors examined
a lattice-fermion hamiltonian that has a ``superfluous" U(1) global symmetry.
They monitored the response of the fermion ground state to an adiabatic
evolution of the (abelian) gauge field that changes the topological charge
of the gauge vacuum.
They found that a U(1) charge of the anticipated amount
is produced in the process.

The clash between chiral-U(1) and gauge
symmetry is at the heart of their argument.
Because of the lack of exact gauge invariance at the lattice level,
the initial and final bare vacua are not gauge-transforms of each other
and their bare U(1) charges are different. During the evolution
the bare charge is necessarily conserved.
But since the bare charge of the ground state changes in the process,
there is a corresponding change in the {\it normal-ordered} charge
defined with respect to the ground state.

The introduction of $m\ne 0$ in our work was necessary to control
the infra-red behavior of a dynamical gauge-fermion system
that undergoes spontaneous symmetry breaking of a peculiar nature.
In contrast, in \rcite{BHS} only the response
of the spectrum of the axial Dirac operator
to an external gauge field was considered,
and so it was not necessary to introduce the mass perturbation.

In conclusion, in this paper we have demonstrated convincingly that,
in spite of the exact chiral U(1) invariance of the lattice action
in the gauge-fixing approach, fermion-number violating processes do
occur, thus resolving the questions raised in \rcite{TB}.

\vspace{5ex}
\noindent {\bf Acknowledgements}

\vspace{2ex}
We would like to thank Pierre van Baal, Aharon Casher
and Mike Creutz for useful discussions.
Both of us would like to thank the Institute for Nuclear Theory at the
University of Washington for hospitality and support provided for
part of this work.
This research is supported in part by a grant from the United-States -- Israel
Binational Science Foundation, and by the US Department of Energy.

\vspace{5ex}
\noindent {\bf Appendix A. Notation}
\secteq{A}
\vspace{3ex}

\noindent The group generators are normalized according to
\beq
  \tr T_a T_b = \half \d_{ab} \,.
\eeq
The dual tensor is
\beq
  \tilde{F}_{\m\n} = \half \e_{\m\n\l\r} F_{\l\r} \,.
\eeq
The topological charge is
\beq
  \n= {g^2 \over 16 \p^2}\; \tr \int d^4x\,F\tilde{F} \,.
\lbl{topo}
\eeq
The hermitian gamma matrices obey the Dirac algebra
\beq
  \{\g_\m \,, \g_\n \} = \d_{\m\n} \,, \qquad
  \g_\m^\dagger = \g_\m \,.
\eeq
In four dimensions we use the representation
\beq
  \g_\m =
  \left(
  \begin{array}{cc}
    0         & \s_\m  \\
    \bar\s_\m & 0
  \end{array}
  \right)
\,,\qquad
  \g_5 =
  \left(
  \begin{array}{cc}
    1  &  0  \\
    0  & -1
  \end{array}
  \right) \,,
\eeq
\beq
  \s_\m=(\vec\s,i) \,, \qquad \bar\s_\m = \s_\m^\dagger =(\vec\s,-i) \,.
\eeq
The chiral projectors are
\beq
  P_R = (1+\g_5)/2 \,, \qquad   P_L = (1-\g_5)/2 \,.
\eeq
Charge conjugation matrices play a key role in the Majorana formulation
of Sect.~3. In any even dimension the charge conjugation matrix is defined by
(see \eg \rcite{AG})
\beq
  C \g_\m = - \g_\m^T C\,,
\lbl{C}
\eeq
and satisfies $C^{-1} = C^\dagger = C^T$.
In $8n+2$ and $8n+4$ dimensions,
$C^T = -C$, while in $8n+6$ and $8n$ dimensions, $C^T = C$.
For the above four-dimensional gamma-matrices the charge-conjugation matrix
can be chosen as
\beq
  C_4 = \g_3 \g_1 =
  \left(
  \begin{array}{cc}
    \e   & 0   \\
    0    & \e
  \end{array}
  \right) \,.
\eeq
It is unique up to a sign.
The two-dimensional anti-symmetric tensor (with $\e_{12} = 1$) is
\beq
  \e = i \s_2 =
  \left(
  \begin{array}{cc}
    0         & 1  \\
    -1        & 0
  \end{array}
  \right) \,.
\lbl{e2dim}
\eeq
For any even dimension and $\m \ne \n$ one has
\beq
  C \g_\m \g_\n = - \g_\m^T C \g_\n = \g_\m^T \g_\n^T C
  = (\g_\n \g_\m)^T C = - (\g_\m \g_\n)^T C \,.
\lbl{Csigma}
\eeq
In four dimensions one has
\beq
  C \g_\m (1 \pm \g_5)  = [ C \g_\m (1 \mp \g_5) ]^T \,,
\eeq
a relation which generalizes to $4n$ dimensions.

\vspace{5ex}
\noindent {\bf Appendix B. SO(10)-ology}
\secteq{B}
\vspace{3ex}

\noindent We define the ten-dimensional gamma matrices by the following
tensor products
\bqry
  \G_1 & = & \s_1 \otimes \s_1 \otimes \s_1 \otimes \s_1 \otimes \s_1 \,, \NON
  \G_2 & = & \s_2 \otimes \s_1 \otimes \s_1 \otimes \s_1 \otimes \s_1 \,, \NON
  \G_3 & = & \s_3 \otimes \s_1 \otimes \s_1 \otimes \s_1 \otimes \s_1 \,, \NON
  \G_4 & = & \idn \otimes \s_2 \otimes \s_1 \otimes \s_1 \otimes \s_1 \,, \NON
  \G_5 & = & \idn \otimes \s_3 \otimes \s_1 \otimes \s_1 \otimes \s_1 \,, \NON
  \G_6 & = & \idn \otimes \idn \otimes \s_2 \otimes \s_1 \otimes \s_1 \,,
\lbl{G10}\NXT
  \G_7 & = & \idn \otimes \idn \otimes \s_3 \otimes \s_1 \otimes \s_1 \,, \NON
  \G_8 & = & \idn \otimes \idn \otimes \idn \otimes \s_2 \otimes \s_1 \,, \NON
  \G_9 & = & \idn \otimes \idn \otimes \idn \otimes \s_3 \otimes \s_1 \,, \NON
\G_{10}& = & \idn \otimes \idn \otimes \idn \otimes \idn \otimes \s_2 \,, \NON
\G_{11}& = & \idn \otimes \idn \otimes \idn \otimes \idn \otimes \s_3 \,.
\nonumber
\eqry
The ten-dimensional charge conjugation matrix is
\beq
  C_{10} = i \, \cc \otimes \s_2 \,,
\lbl{C10}
\eeq
where
\beq
  \cc =  \s_2  \otimes \s_3 \otimes \s_2 \otimes \s_3 \,.
\lbl{cc}
\eeq
Notice that $C_{10}$ anti-commutes with $\G_{11}$.
With the above definitions, \seeq{Csigma} reads
\beq
  \left(
  \begin{array}{cc}
    0         & \cc  \\
    -\cc      & 0
  \end{array}
  \right)
  \left(
  \begin{array}{cc}
    \S_{\ssstyle MN} & 0  \\
    0                & \bar\S_{\ssstyle MN}
  \end{array}
  \right)
  = -
  \left(
  \begin{array}{cc}
    \S_{\ssstyle MN}^T & 0  \\
    0                  & \bar\S_{\ssstyle MN}^T
  \end{array}
  \right)
  \left(
  \begin{array}{cc}
    0         & \cc  \\
    -\cc      & 0
  \end{array}
  \right) \,.
\eeq
Equivalently
\beq
  \cc \, \bar\S_{\ssstyle MN} = - \S_{\ssstyle MN}^T \, \cc
   = - \S_{\ssstyle MN}^* \, \cc \,,
\lbl{CSig}
\eeq
a relation which is needed for the derivation of \seeq{majrel}.

If $D$ satisfies \seeq{majrel} and has no zero modes,
its inverse satisfies
\beq
  G(x,y) \equiv \svev{\J(x)\,\bJ(y)}
  = - C_4 \cc \, G^T(y,x) \, C_4 \cc \,.
\eeq
By taking a suitable limit, this generalizes to the case where
$D$ has exact zero modes and $G(x,y)$ is constructed from the non-zero
modes only.

We now show that, for almost every embedding of an instanton in SO(10),
$\det(\l) \ne 0$ ({cf.} \seeq{lt}),
\ie the mass term of Sect.~3 lifts the four zero modes.
We will first show that $\det(\l) \ne 0$ for a particular embedding.
We introduce 16 by 16 matrices $\cs_i(l)$, $i=1,\ldots,3$, $l=1,\ldots,4$,
defined to be tensor products of four two-by-two matrices
with $\s_i$ as the $l$-th factor and the identity for the rest.
The SO(10) generators $\S_{\ssstyle MN}$ with
${\sstyle M},{\sstyle N}=1,\ldots,4$, generate two SU(2) groups.
For $\cs_3(2) = \pm 1$ we label them SU(2)$_{R,L}$.

Each zero mode is written explicitly as $u_i=u_{\a\b_1\b_2\b_3\b_4,i}(x)$,
where $\a,\b_1,\b_2,\b_3,\b_4=1,2.$ Here $\a$ is the spin index.
The other four indices correspond to the tensor product that defines the SO(10)
generators in the {\sixteen} representation.
Assuming that the instanton resides
in \eg the SU(2)$_L$ subgroup defined above,
the zero modes have $\cs_3(2) = -1$ (equivalently $\b_2=2$).
Their SU(2)$_L$ index is $\b_1$.
Using the explicit solution of the isospin one-half zero mode
(for a regular-gauge instanton) we have
\beq
  u_{\a\b_1\b_2\b_3\b_4,i}(x)
  = \cn\,\d_{\a,\b_1} \d_{\b_2,2}\, \co_{\b_3\b_4,i}\, f(x^2) \,,
\lbl{bbbb}
\eeq
where $f(x^2)=(x^2+\r^2)^{-3/2}$ and $\cn$ is a normalization factor.
The constants $\co_{\b_3\b_4,i}$ define the four independent zero modes.
We will label them by the eigenvalues of $\cs_2(3)$ and $\cs_3(4)$.
Replacing the index $i=1,\ldots,4,$ by a pair of indices $\t_1,\t_2=1,2$,
we take $\co_{\b_3\b_4,\t_1\t_2}=i(\s_2)_{\b_3\t_1} (\s_3)_{\b_4\t_2}$.
In matrix notation, $\co=i\s_2\otimes\s_3$, and $\co$
has similar properties to the four-dimensional charge-conjugation matrix.
Putting together \seeqs{lt}, \seneq{cc} and\seneq{bbbb} we get
\bqry
  \l_{\t_1\t_2,\t'_1\t'_2}
  & = &
  -m\, \cn^2 \int d^4x\, f^2(x^2)\; \tr(\e^2)
  \left(\co^T (i\s_2\otimes\s_3) \, \co \right)_{\t_1\t_2,\t'_1\t'_2}
\NON
   & = & m\, (i\s_2)_{\t_1\t'_1}(\s_3)_{\t_2\t'_2}\,,
\lbl{cu}
\eqry
which proves $\det(\l) \ne 0$ for this special case. On the first row,
the explicit $i\s_2\otimes\s_3$ comes from the last two factors
in the tensor product\seneq{cc}, while $\tr \e^2$ comes from
the first factor in this tensor product, and the explicit $\e$ in \seeq{lt}.
The transition from the first to the second row implicitly defines
the normalization constant.

Suppose now that a global rotation $R\!\in\,$SO(10) is applied to
the above special embedding of the instanton. The new zero modes are
$u'_i = R\, u_i$. We claim that $\det(\l(R)) \ne 0$
for almost every $R$. The proof is simple. Suppose on the contrary
that $\det(\l(R)) = 0$ for every $R$ in some open subset of SO(10).
Since the embedding and, hence, $\det(\l)$ are analytic functions
of $R$, this would imply that $\det(\l)=0$ for all $R$.
This, however, contradicts \seeq{cu} in the special case $R=I$.
Therefore $\det(\l)=0$ may be true, at most,
on a measure zero subset of SO(10).

\vspace{5ex}
\noindent {\bf Appendix C. Lattice formulae}
\secteq{C}
\vspace{3ex}

The free symmetric lattice derivative is defined for any function $f_x$ by
\bqry
  \hp_{x,\m} f = {1\over 2a} (f_{x+\hat\m} - f_{x-\hat\m}) \,,
\nonumber
\eqry
where $\hat\m$ is a unit vector in the $\m$-direction.
The corresponding covariant derivative is
\bqry
  \hd_{x,\m} f
  = {1\over 2a}
  (U_{x,\m}\, f_{x+\hat\m} - U^\dagger_{x-\hat\m,\m}\, f_{x-\hat\m}) \,,
\nonumber
\eqry
where $U_{x,\m}$ is the link variable. The free lattice laplacian is
\bqry
  \hb_{x,\m} f = {1\over a^2}
                 \left(f_{x+\hat\m} + f_{x-\hat\m} - 2 f_{x}\right) \,.
\nonumber
\eqry
Given a set of left-handed fields $\j_L^i$ and corresponding spectator
fields $\c_R^i$, the chiral Wilson lagrangian is
(compare the upper-right block in \seeq{cW})
\beq
  \cl = \sum_i \left(
      \bji_L \,\s_\m\hd_\m\, \j_L^i + \bci_R \,\bar\s_\m\hp_\m\, \c_R^i
      -{ar\over 2} (\bci_R \,\hb\, \j_L^i + \bji_L \,\hb\, \c_R^i)
      \right)\,,
\lbl{cwl}
\eeq
where $r$ is Wilson parameter.
This action is invariant under a U(1) rotation of all fermion fields
\beq
  \j_L^i \to e^{i\a} \j_L^i\,, \quad \c_R^i \to e^{i\a} \c_R^i\,,\quad
  \bji_L \to e^{-i\a}\, \bji_L\,, \quad \bci_R \to e^{-i\a}\, \bci_R\,.
\lbl{U1}
\eeq
The conserved U(1) current is (see \rcite{pt})
\bqry
  \jllt_{x,\m}
&=& {1\over 2} \sum_i \Bigg\{
                \bji_{L,x}\s_\m U_{x,\m}\j^i_{L,x+\hat\m}
                +\bji_{L,x+\hat\m} \s_\m U_{x,\m}^\dagger \j^i_{L,x}
\NON
&&              +\bci_{R,x}\bar\s_\m \c^i_{R,x+\hat\m}
                +\bci_{R,x+\hat\m} \bar\s_\m \c^i_{R,x}
\lbl{cWJ}
\NXT
&&    -r\Big(\bji_{L,x} \c^i_{R,x+\hat\m} + \bci_{R,x} \j^i_{L,x+\hat\m}
      -\bji_{L,x+\hat\m} \c^i_{R,x} - \bci_{R,x+\hat\m} \j^i_{L,x} \Big)
  \Bigg\} \,.
\nonumber
\eqry
It satisfies the conservation equation
\beq
  \sum_\m (\jllt_{x,\m} - \jllt_{x-\hat\m,\m}) = 0 \,.
\lbl{Jcons}
\eeq
In the special case of one-flavor QCD let us introduce Dirac fermions
$\j=(\j_R,\j_L)$, $\bj=(\bj_L,\bj_R)$, $\c=(\c_R,\c_L)$, $\bc=(\bc_L,\bc_R)$.
The axial transformation is
\beq
  \j \to e^{-i\a\g_5} \j\,, \quad \c \to e^{i\a\g_5} \c\,,\quad
  \bj \to \bj\, e^{-i\a\g_5}\,, \quad \bc \to  \bc\, e^{i\a\g_5}\,,
\eeq
and the axial current is
(note that in \seeq{cW} we set $r=1$)
\bqry
  \jlt_{x,5\m}
&=& {1\over 2} \Bigg\{
                \bj_{x}\g_5\g_\m U_{x,\m}\j_{x+\hat\m}
                +\bj_{x+\hat\m} \g_5\g_\m U_{x,\m}^\dagger \j_{x}
\NON
&&              -\bc_{x}\g_5\g_\m \c_{x+\hat\m}
                -\bc_{x+\hat\m} \g_5\g_\m \c_{x}
\lbl{axWJ}
\NXT
&&        -r\Big(\bj_{x}\g_5\c_{x+\hat\m} - \bc_{x}\g_5\j_{x+\hat\m}
                -\bj_{x+\hat\m}\g_5\c_{x} + \bc_{x+\hat\m}\g_5\j_{x} \Big)
  \Bigg\} \,.
\nonumber
\eqry
For $m\ne 0$ the axial current satisfies the partial-conservation equation
\beq
  {1\over a} \sum_\m (\jlt_{x,5\m} - \jlt_{x-\hat\m,5\m})
  = -2m J_{x5}\,.
\lbl{parcons}
\eeq
The difference operator on the left-hand side (the free backward derivative)
becomes $(1-\exp(-iap_\m))/a = ip_\m + \ldots$, in momentum space.
The local scalar and pseudo-scalar lattice densities are
$\bj_x \j_x$ and  $J_{x5} = \bj_x \g_5 \j_x$.
As usual they are related by an axial rotation.
They do not mix with the corresponding spectator-field densities
thanks to the shift symmetry~\cite{GP}.

We now explain why singular-gauge instantons are suppressed on
the lattice by the gauge-fixing action of \rcite{gfx}.
Recall that, in a singular gauge,  the instanton's vector potential
near the gauge singularity (located at $x=0$) is
$A_\m \sim \F(x) \pd_\m \F^\dagger(x)/g$ where $\F(x) = \s_\m x_\m / |x|$.
The magnitude of this vector potential grows like $1/(g|x|)$.

On the lattice let us make the (bare-field) rescaling
$A_{x,\m} \to A'_{x,\m}=g_0 A_{x,\m}$. Dropping the prime,
we expand the link variable as $U_{x,\m}=\exp(iaA_{x,\m})$.
The gauge-fixing action contains the expected longitudinal
kinetic term
\beq
  {1\over \x_0 g_0^2}\; \tr\left(\sum_\m \pd_\m A_\m \right)^2
  = {1\over 2\x_0 g_0^2}\; \sum_a \left(\sum_\m \pd_\m A_\m^a \right)^2\,,
\eeq
plus irrelevant terms. Here $\x_0$ is the bare gauge-fixing parameter
and index summations have been shown explicitly.
The leading irrelevant term that contributes to the classical potential is
\beq
  {a^2\over 2\x_0 g_0^2}\; \tr\left(\sum_\m A_\m^2 \; \sum_\n A_\n^4\right)\,.
\lbl{A6}
\eeq
The irrelevant terms break BRST invariance, and so there is no reason
that regular-gauge and singular-gauge instantons will have the same
lattice action.

Consider now some lattice discretization of the singular-gauge
vector potential. Inevitably, the (rescaled) vector potential will be $O(1/a)$
in the hypercube(s) containing the point $x=0$ and the vicinity.
For such a vector potential $U_{x,\m}-I = O(1)$.
The positivity of expression\seneq{A6} and
(since there are infinitely-many other irrelevant terms)
of the gauge-fixing action as a whole~\cite{gfx},
guarantees that the lattice action will be an $O(1)$ quantity {\it times}
$1/g_0^2$. In the continuum limit $g_0 \to 0$ any such lattice configuration
is suppressed.

\vspace{5ex}
\noindent {\bf D. Continuum propagators in the presence of approximate
zero modes}
\secteq{D}
\vspace{3ex}

Let $H=H_0 +\a V$ be a Schr\"odinger-like (\ie elliptic and self-adjoint)
operator in a $d$-dimensional open infinite space.
We assume that $H_0$ has a finite number
of (normalized) zero modes $u_{i}(x)$, and that $H$ has no zero modes.
The inverse of $H$, denoted $G$, is defined by
\beq
  H G = 1 \,,
\lbl{G}
\eeq
where both sides are considered as operators acting on a suitable
Hilbert space. In this appendix we explain how to construct
a systematic approximation for $G$.
(\seEq{G} may be rewritten in the familiar form
$H G(x,y) = \d(x-y)$ by taking the matrix element of \seeq{G}
between ``position eigenstates'' $\sbra{x}$ and $\sket{y}$.)

Let us introduce the notation $\sket{i}$ for the zero modes of $H_0$,
and $\sket{p}$ for the rest of its spectrum.
(In instanton problems the zero modes are the only bound states,
and $\sket{p}$ denotes the continuum of scattering states.)
One has $H_0\sket{p} = E(p)\sket{p}$. The propagator $G_0$ is defined by
\beq
  H_0 G_0 = 1 -  \sket{i}\sbra{i} = \sket{p}\sbra{p}\,,
\lbl{G0}
\eeq
and has the spectral representation
\beq
  G_0 = \sket{p} (G_0)_{pq} \sbra{q} \,, \qquad\qquad
  (G_0)_{pq} = E^{-1}(p)\,\d(p-q) \,.
\lbl{G0rep}
\eeq
It is convenient to expand $G=G(\a)$ too using the eigenmodes of $H_0$
as a complete orthonormal basis
\beq
  G = \sket{i} \b^{-1}_{ij} \sbra{j}
      \, + \, \sket{p} G_{pq} \sbra{q}
      \, + \, \sket{p} f_{pj} \sbra{j}
      \, + \, \sket{j} f^*_{pj} \sbra{p} \,.
\lbl{expd}
\eeq
In this expansion, the basis vectors are fixed ($\a$-independent),
while the $\a$-dependence is carried by the spectral functions
$\b_{ij}$, $G_{pq}$ and $f_{pj}$.
Below, we show that the spectral functions can be expanded as power series
in $\a$ where $\b_{ij}$ is $O(\a)$, and $G_{pq}$ and $f_{pj}$ are $O(1)$.
They will be used to construct approximations for $G$.

We start by substituting \seeq{expd} into \seeq{G}.
Using $H_0 \sket{i}=0$ we get
\beq
  1 = \a V \sket{i} \b^{-1}_{ij} \sbra{j} + H G' \,,
\lbl{subt}
\eeq
where $G'$ consists of the last three terms on the right-hand side
of \seeq{expd}. By taking the matrix element between zero-mode states
$\sbra{i}$ and $\sket{j}$ and using $\langle p|i\rangle=0$ we get
\beq
  \d_{ij} = \a \sbra{i} V \sket{k} \b^{-1}_{kj}
            + \a \sbra{i} V \sket{q} f_{qj} \,.
\lbl{ij}
\eeq
By taking the matrix element between $\sbra{p}$ and $\sket{j}$ we get
\beq
  0 = \a \sbra{p} V \sket{k} \b^{-1}_{kj}
      + H_{pq} f_{qj} \,,
\lbl{pj}
\eeq
where
\beq
  H_{pq} = \sbra{p} H \sket{q} \,.
\lbl{pq}
\eeq
In order to solve for $\b_{ij}$ and $f_{qj}$ we have to invert $H_{pq}$.
Note that $H_{pq}$ is the continuous-index kernel
of the operator $H^\perp \equiv \sket{p} H_{pq} \sbra{q}$
which is, by definition, the projection of $H$ onto the subspace orthogonal
to the zero modes of $H_0$. The inverse of the projected operator is
defined by $H^\perp\, (H^\perp)^{-1} =  \sket{p} \sbra{p}$.
The corresponding kernel satisfies
$(H^\perp)^{-1} \equiv \sket{p} H_{pq}^{-1} \sbra{q}$.
One has
\beq
  \sbra{p} H \sket{q} = \sbra{p} H_0 \sket{q} + O(\a)
  = E(p)\, \delta(p-q) + O(\a) \,.
\eeq
Therefore $(H^\perp)^{-1}$ exists, and
(compare \seeq{G0rep})
\beq
  H_{pq}^{-1} = E^{-1}(p)\, \delta(p-q) + O(\a) \,.
\lbl{Hinv}
\eeq

We are now ready to solve for $\b_{ij}$ and $f_{qj}$.
Multiplying \seeq{pj} by $H_{pq}^{-1}$ we get
\beq
  f_{qj} = - \a H^{-1}_{qp} \sbra{p} V \sket{k} \b^{-1}_{kj} \,,
\lbl{f}
\eeq
and substituting this in \seeq{ij} we get
\beq
  \b_{ij} = \left\langle i \left| \; \left(
            \a V - \a^2 V \sket{q} H^{-1}_{qp} \sbra{p} V
            \right)\; \right| j \right\rangle \,.
\lbl{beta}
\eeq
\seEq{Hinv} implies that $H^{-1}_{qp}=O(1)$,
hence $\b_{ij} = O(\a)$ and $f_{qj}=O(1)$.

The parametric $\a$-dependence of the last spectral function is determined
to be $G_{pq}=O(1)$. To show this, take the matrix element
of \seeq{subt} between scattering states $\sbra{p}$ and $\sket{q}$,
which gives
\beq
  \d(p-q) = H_{pp'} G_{p'q} + \a \sbra{p} V \sket{k} f^*_{qk} \,.
\eeq
Since we already know that $f_{qj}=O(1)$, the last term on the right-hand
side is sub-leading. It follows that
$G_{pq} = H_{pq}^{-1} + O(\a) = (G_0)_{pq} + O(\a)$.
Moreover, by combining these results it follows that
the spectral functions can be expanded as power series in $\a$,
starting at the above-specified order for each spectral function.
(See, however, {\it Comment 1} below.)

The propagator $G$ involves $\b_{ij}^{-1}$,
and so it has a Laurent series starting at order $1/\a$.
To find the singular, $O(1/\a)$ piece of the propagator,
we keep only the $O(\a)$ term on the right-hand side of \seeq{beta}.
We get $\b_{ij}=\l_{ij,1}+O(\a^2)$,
where
\beq
  \l_{ij,1} = \a \sbra{i} V \sket{j} \,.
\lbl{l1}
\eeq
This expression is recognized as the first-order energy shifts
of the zero modes, as calculated using degenerate perturbation theory.
Substituting in \seeq{expd} we obtain
\beq
  G = \sket{i} \l^{-1}_{ij,1} \sbra{j} + O(1) \,.
\lbl{Gx0}
\eeq

While \seeq{Gx0} is all we will be using in the body of the paper,
it is instructive to go one step further and construct
also the $O(1)$ part of the propagator.
To this end we approximate $H_{pq}^{-1}$ by $(G_0)_{pq}$.
\seEq{beta} then yields $\b_{ij}=\l_{ij,2}+O(\a^3)$ where
\beq
  \l_{ij,2} = \sbra{i} \a\, V
                -\a^2\,  V\, G_0\, V \sket{j} \,,
\lbl{l2}
\eeq
which, as expected, includes also the second-order energy shifts.
Next, making a similar approximation in \seeq{f} we have
\beq
  \sket{p} f_{pj} \sbra{j}
  = - \a\, G_0 V \sket{i} \l^{-1}_{ij,1} \sbra{j} + O(\a) \,.
\eeq
This involves the first-order correction to the zero mode's wave function.
Putting everything together we find that, at $O(1)$, the propagator
may be compactly expressed as
\beq
  G = \sket{i,1} \l^{-1}_{ij,2} \sbra{j,1}  + G_0 + O(\a) \,,
\lbl{Gx1}
\eeq
where
\beq
  \sket{i,1} = \left( 1 - \a G_0 V \right) \sket{i} \,.
\eeq

\noindent We conclude this appendix with two technical comments.

\vspace{1ex}
\noindent {\it Comment 1.}
Perturbation theory is {\it a-priori} not valid,
if the perturbation $\a V$ results in the disappearance of any bound states
from the spectrum.
What we have done amounts to showing that perturbation
theory can still be used to construct an $O(\a^n)$ approximation
for the propagator, provided that the integrals occurring at
$(n+1)$-th order in perturbation theory converge.
As already mentioned, in this paper we actually use only \seeq{Gx0}.
In four dimensions fermionic zero modes fall like $|x|^{-3}$ or faster
for $|x| \to \infty$. Even if we perturb by a spatially constant mass term $m$
(that does not vanish at infinity) \seeq{Gx0} gives the correct expression
for the $O(1/m)$ part of the propagator.

\vspace{1ex}
\noindent {\it Comment 2.}
In a semi-classical calculation we also have to separate out
the approximate zero modes' contribution to the determinant.
As discussed in detail in \rcite{susy},
this can be done by splitting the functional integration
into separate integrations over the amplitudes of the (approximate) zero modes,
and over the orthogonal subspace.
To leading order, the integration over the zero modes' subspace
gives rise to $\det(\l_1)$ (see \seeq{l1} above).
The integration over the orthogonal subspace
has a non-vanishing finite limit (after subtracting UV divergences).

\vspace{5ex}

\end{document}